\documentclass[11pt,a4paper]{article}
\usepackage{graphics}
\begin{document}
\title{\textbf{On a possible EPR experiment with \boldmath{$B^{0}_{d}\overline{B^{0}_{d}}$} pairs}\vspace{2cm}} 
\author{A. Pompili \\ \small{Dipartimento di Fisica, Universit\`a di Trieste and INFN Trieste,} \\  \small{Area di Ricerca, Padriciano 99, I-34012 Trieste, Italia} \vspace{0.5cm}\\ F. Selleri \\  \small{Dipartimento di Fisica, Universit\`a di Bari and INFN Bari,} \\  \small{via Amendola 173, I-70126 Bari, Italia}}

\date{}
\maketitle
\vspace{2cm}
\begin{abstract}
A very general local realistic theory of single $B^{0}_{d}$ mesons and of correlated $B^{0}_{d}\overline{B^{0}_{d}}$ pairs is formulated. If these pairs are produced in the $\Upsilon (4S)$ decay, the local realistic asymmetry for observing pairs with like and unlike flavour at different proper times remarkably differs from the quantum mechanical prediction. Asymmetric B-factories provide a powerful tool for the study of the EPR problem since the relative detection experiments are shown to be capable of a time-dependent measurement precise enough to discriminate between local realism and quantum theory.
\medskip
\noindent
\end{abstract}
\vspace{2.7cm}

\pagebreak
\section{INTRODUCTION}
\indent

The strange nature of quantum correlations between separated systems, pointed out for the first time by Einstein, Podolsky and Rosen (EPR) [1], has stimulated a lively debate. The incompatibility between quantum theory predictions and local realism (LR) consequences became evident with the 1965 work of Bell [2], showing that a wide class of local hidden-variable models satisfies an inequality which is violated by quantum mechanics (QM).

In 1969 Clauser, Holt, Shimony and Horne [3] stressed that Bell's inequality could be experimentally checked with photon pairs emitted by single atoms, even with the available low efficiency photon counters, if a suitable additional assumption was made. Several experimental investigations of the EPR paradox have accordingly been performed, mostly with photon-polarization correlation measurements using radiative atomic cascade transitions [4]. In these experiments the inequality was found to be violated and the quantum mechanical predictions turned out to agree with the data. It has been pointed out, however, that the introduction of additional assumptions has led to the formulation of inequalities different from (and stronger than) Bell's original inequality [5,6]. The experimental results violated the stronger inequalities but were still compatible with Bell's original one, which was deduced from local realism alone. This allows the argument that the evidence is against the additional assumptions but not against LR. Therefore from a strictly logical point of view, the choice between LR and the existing quantum theory has to be done yet. 

Since the atomic physics experiments are not loophole free and thus have not been able to settle unambiguously that LR should be discarded, it is worthwhile to study the EPR paradox in domains where highly efficient particle detectors can be used and the additional assumptions are not needed. Indeed the behaviour of a pair of neutral pseudo-scalar mesons (e.g. $K^{0}\overline{K^{0}}$ or $B^{0}\overline{B^{0}}$), anti-correlated in flavour if produced by the decay of a $J^{PC} \! = \! 1^{--}$ state (e.g. the $\Phi$ or the $\Upsilon(4S)$ resonances) seems even more puzzling than the behaviour of spin correlated pairs of stable particles (photons or electrons).

A critical discussion of the EPR paradox for $K^{0}\overline{K^{0}}$ pairs and of the earlier attempts to check Bell's inequality in kaon $\Phi$ decays was made by Ghirardi, Grassi and Weber [7]. The main argument was that Bell's inequality, written in terms of four different times of flight of the kaons, is not violated by the quantum-mechanical two-time joint probability for correlated strangeness, due to the specific values of kaon masses and decay widths. It could easily be shown in the same way that an analogous argument holds for $B_{d}$ mesons.

Recently further Bell-type tests involving new Bell-like inequalities for correlated neutral meson-anti meson pairs have been proposed and discussed [8]. However, some of them would probe only a restricted class of local realistic theories whereas others avoid this difficulty but require experimental set-ups not available in the near future [6].

But Bell's inequality is only one of the many consequences of LR: EPR correlations can provide tests sensitive to possible deviations from QM. Meaningful tests of LR, not of the Bell-type, have been proposed for the $K^{0} \! - \! \overline{K^{0}}$ system [9,10]. In the present paper, after developing a very general local realistic theory for the $B^{0}_{d} \! - \! \overline{B^{0}_{d}}$ system (following the ideas in [9] and [10]), a test to discriminate between QM and LR is proposed and its feasibility with an asymmetric B-factory is discussed.

Only a few works on EPR tests refer to the $B^{0}_{d}\! - \! \overline{B^{0}_{d}}$ system [11] and use a \textit{decoherence parameter} $\zeta$ (such that the QM interference term is multiplied by a factor $(1 \! - \! \zeta)$) to measure deviations from QM. Combining already presently data (from \textit{CLEO} and \textit{Argus} and from LEP experiments) they deduce a clear indication in favour of QM. However this is an expected conclusion taking into account that falsification of $\zeta \! = \! 1$ means simply a falsification of the spontaneous factorization hypothesis (SFH). Indeed SFH was already falsified with gamma ray pairs from $e^{-}\! - e^{+}$ annihilations [12], with atomic photon pairs [13] and recently with kaons pairs produced in $p \! - \! \overline{p}$ annihilations [14]. On the other hand it is possible to reproduce within the local realistic approach all non paradoxical predictions of QM, anti-correlations in \textit{strangeness} and \textit{CP} (or \textit{beauty} and \textit{mass}) included, which are absent in SFH; thus SFH violation is also predicted by any reasonable approach based on LR.

This work is organized as follows. In the second section the B-meson pair correlation and the quantum-mechanical formalism are reviewed. In the third section a local realistic theory for the $B^{0}_{d} \! - \! \overline{B^{0}_{d}}$ system is accurately developed and its predictions, different from those of QM, are discussed. In the fourth section we stress the asymmetric collider allows to perform time-dependent measurement and then discuss how our EPR test can be performed using techniques and methods peculiar to lifetime and mixing measurement.

\section{B-MESON PAIR CORRELATION}
\indent

Let us briefly review the quantum-mechanical basic formalism for the EPR-correlated $B^{0}\overline{B^{0}}$ pairs that can be created as decay products of the $\Upsilon(4S)$ resonance. More precisely $B^{0}_{d}\overline{B^{0}_{d}}$ pairs are produced since $\Upsilon(4S)$ is not heavy enough to decay into $B^{0}_{s}\overline{B^{0}_{s}}$ pairs. From now on the index \textit{d} will be dropped to simplify notation: with $B^{0}$ a $B^{0}_{d}$ will be intended. 

In the quark model the resonance $\Upsilon(4S)$ is a $b\overline{b}$ state with quantum numbers $J^{PC} \! = \! 1^{--}$. In the strong decay $\Upsilon(4S) \! \rightarrow \! B^{0}\overline{B^{0}}$ the created pair inherits the $\Upsilon(4S)$ quantum numbers. Since B-mesons are spinless $J^{P} \! = \! 1^{-}$ implies that the $B^{0}\overline{B^{0}}$ pair is in a \textit{p-wave} total angular momentum state. $C \! = \! -1$ implies that the flavour part of the pair wavefunction is antisymmetric. Therefore immediately after the decay (namely at $t \! = \! 0$) the $B^{0}\overline{B^{0}}$ state vector is given, in the $\Upsilon(4S)$-frame, by 

\begin{eqnarray} \begin{array}{ll} |\psi(0,0) \! >= \! \frac{1}{\sqrt{2}} \! \{ |B^{0}(\vec{p}) \! \! >|\overline{B^{0}}(-\vec{p}) \! >-|\overline{B^{0}}(\vec{p}) \! >|B^{0}(-\vec{p}) \! \! > \} \!\! \equiv \! \frac{1}{\sqrt{2}} \!\! \{|B^{0} \! \! >_{l}|\overline{B^{0}} \!\! >_{r}-|\overline{B^{0}} \!\! >_{l}|B^{0} \!\! >_{r} \} \end{array} \end{eqnarray}
where $l$($r$) denote the B-mesons motion directions ``left''(``right'') and $|B^{0}>$, $|\overline{B^{0}}>$ are \textit{beauty eigenstates} ($B^{0} \! = \! \overline{b}d$, $\overline{B^{0}} \! = \! b\overline{d}$).

Possible tiny CP-violation will be neglected throughout this paper since it could not appreciably modify the large difference between LR and QM predictions that will be found. Thus the \textit{mass eigenstates} $|B^{0}_{H}>$, $|B^{0}_{L}>$ ($H$ for ``heavy'' and $L$ for ``light'' such that $\Delta m \! = \! m_{H} \! - \! m_{L} \! > \! 0$) can be identified with \textit{CP eigenstates}:

\begin{equation} |B^{0}_{H}>=\frac{1}{\sqrt{2}} \{ |B^{0}>+|\overline{B^{0}}> \} \hspace{1.0cm},\hspace{1.0cm} |B^{0}_{L}>=\frac{1}{\sqrt{2}} \{ |B^{0}>-|\overline{B^{0}}> \} \end{equation}
and the state vector (1) can also be expressed as

\begin{equation} |\psi(0,0)>=\frac{1}{\sqrt{2}} \{ |B^{0}_{H}>_{l}|B^{0}_{L}>_{r}-|B^{0}_{L}>_{l}|B^{0}_{H}>_{r} \} \end{equation}

The time evolution of the complex mass eigenstates is given by

\begin{equation} |B^{0}_{H}(t)>=|B^{0}_{H}>e^{-\alpha_{H}t} \hspace{1cm},\hspace{1cm} |B^{0}_{L}(t)>=|B^{0}_{L}>e^{-\alpha_{L}t} \end{equation}
where $t$ is the particle proper time and (with $\hbar = c = 1$)

\begin{equation} \alpha_{H}=\frac{1}{2}\Gamma+im_{H} \hspace{1cm},\hspace{1cm} \alpha_{L}=\frac{1}{2}\Gamma+im_{L} \end{equation}
(the two neutral B-mesons are expected to have a negligible difference in lifetime).

The time evolution operator for state (1) is the product of the time evolution operators for single B-mesons states, so that, at proper times $t_{l}$ and $t_{r}$, the time evolved state for the $B^{0}\overline{B^{0}}$ pair can be written as

\begin{equation} |\psi(t_{l},t_{r})>=\frac{1}{\sqrt{2}} \{ |B^{0}_{H}>_{l}|B^{0}_{L}>_{r}e^{-\alpha_{H}t_{l}-\alpha_{L}t_{r}} - |B^{0}_{L}>_{l}|B^{0}_{H}>_{r} e^{-\alpha_{L}t_{l}-\alpha_{H}t_{r}} \} \end{equation}

The different exponentials in (6) generate $B^{0}{B}^{0}$ and $\overline{B^{0}}\overline{B^{0}}$ components. Indeed the time evolution of $B^{0}$, $\overline{B^{0}}$ is governed by a weak interaction that does not conserve flavour and thus allows $B^{0} \! - \! \overline{B^{0}}$ oscillations to take place between the \textit{flavour eigenstates} ($B^{0} \! - \! \overline{B^{0}}$ \textit{mixing}). The probabilities of $B^{0}\overline{B^{0}}$, $B^{0}{B}^{0}$ and $\overline{B^{0}}\overline{B^{0}}$ observations are given (using (6) and (2)) by

\begin{equation} P^{QM}[B^{0}(t_{l});\overline{B^{0}}(t_{r})]=\frac{1}{4}E(t_{r}+t_{l})[1+cos( \Delta m (t_{r}-t_{l}))]=P^{QM}[\overline{B^{0}}(t_{l});B^{0}(t_{r})] \end{equation}

\begin{equation} P^{QM}[B^{0}(t_{l});B^{0}(t_{r})]=\frac{1}{4}E(t_{r}+t_{l})[1-cos(\Delta m (t_{r}-t_{l}))]=P^{QM}[\overline{B^{0}}(t_{l});\overline{B^{0}}(t_{r})] \end{equation}
where $E(t) \! \equiv \! e^{-\Gamma t}$.

For $t_{l} \! = \! t_{r}$ eq. (8) vanishes and the probability of having like flavours is zero. This flavour anti-correlation means that the two neutral B-mesons evolve in phase so that at equal proper times, until one of them decays, there are always a $B^{0}$ and a $\overline{B^{0}}$ present. 

Let us consider the following flavour asymmetry

\begin{equation} A(t_{l},t_{r}) = \frac{ P[B^{0}(t_{l}); \overline{B^{0}}(t_{r})]-P[\overline{B^{0}}(t_{l}); \overline{B^{0}}(t_{r})] } { P[B^{0}(t_{l});\overline{B^{0}}(t_{r})]+P[\overline{B^{0}}(t_{l});\overline{B^{0}}(t_{r})] }\end{equation}
that will be our fundamental parameter for the comparison between quantum theory and the local realistic predictions. From the experimental point of view the advantage of considering ratios of probabilities or number of events is that systematic errors tend to cancel.

In QM the asymmetry (9) is predicted to be a very simple function of $t_{r} \! - \! t_{l}$ only (by (7) and (8)):

\begin{equation} A^{QM}(t_{l},t_{r})=cos(\Delta m(t_{r}-t_{l}))=cos(x(\frac{t_{r}-t_{l}}{\tau})) \end{equation}
where [15] 

\begin{equation} x=\frac{\Delta m}{\Gamma}= 0.723\pm 0.032 \end{equation} 

We will show that the asymmetry predicted by LR is necessarily quite different from (10).

\section{B-MESON PAIR CORRELATIONS WITHIN LR}
\subsection{Two time probabilities}
\indent

The EPR paradox arises from the incompatibility at the empirical level between the predictions of quantum theory and local realism. The latter can be expressed by the following three assumptions:  
               
(I) If, without in any way disturbing a system, we can predict with certainty the value of a physical quantity, then there exists an element of physical reality corresponding to this physical quantity (\textit{EPR reality criterion}).  

(II) If two physical systems (e.g. our two neutral B mesons) are separated by a large distance, an element of reality belonging to one of them cannot have been created by a measurement performed on the other one (\textit{separability}).
        
(III) If at a given time $t$ a physical system has an element of reality, the latter cannot be created by measurements performed on the same system at time $t' \! > \! t$ (\textit{no retroactive causality}).

Local realism can be applied to the $B^{0}$ meson pairs described quantum mechanically by the state vector (6), by considering only those predictions of (6) to which EPR reality criterion can be applied. These are \textit{flavour} and \textit{mass} anti-correlations. Obviously, if such anti-correlations were not found to exist experimentally, our conclusions could not be correct. If they are assumed to exist, one can say the following [9]:

a) each B meson of every pair has an associated element of reality $\lambda_{1}$ which determines a well defined value of mass ($\lambda_{1} \! = \! +1, \! -1$ corresponds to $m_{H},m_{L}$, respectively);
                                
b) each B meson of every pair has an associated element of reality $\lambda_{2}$ which determines a well defined value of flavour ($\lambda_{2} \! = \! +1, \! -1$ corresponds to $b \! = \! +1, \! -1$, respectively). Furthermore $b$ is not a stable property: it has sudden jumps, from $+1$ to $-1$ and vice-versa, that are simultaneous for the two mesons of every pair but happen at random times in a statistical ensemble of many pairs. 

Notice that the application of local realism to the physical situation described by (6) has brought us, at least formally, outside quantum theory: no quantum mechanical state vector exists, in fact, which can describe a B meson as having simultaneously well defined mass and flavour values. This is of course the standard approach of all realistic interpretations of quantum phenomena.

Following the treatment in [9] and [10] relative to K-mesons and taking into account the analogy (\textit{strangeness},\textit{CP}) $\leftrightarrow$ (\textit{beauty}, \textit{mass}), four B-meson basic states, characterized by \textit{beauty} and \textit{mass} both definite, can be introduced:

\begin{eqnarray} \begin{array}{llll} B_{1}=B_{H} \; \; : \; \; state \; \; with\; \; b \! = \! +1 \; \; and \; \; m \! = \! m_{H} \\ B_{2}=\overline{B_{H}} \; \; : \; \; state\; \; with \;  \; b \! = \! -1 \; \; and \; \; m \! = \! m_{H} \\ B_{3}=B_{L} \; \; : \; \; state \; \; with \; \; b\! = \! +1 \;\; and \; \; m \! = \! m_{L} \\B_{4}=\overline{B_{L}} \; \; : \; \; state \; \; with \; \; b \! = \! -1 \; \; and \; \; m \! = \! m_{L} \end{array} \end{eqnarray}
The probabilities of observing the i-th state at proper time t conditional on an initial j-th state at proper time zero are:

\begin{center} $p_{ij}(t|0) =$ probability of $B_{i}$ at time $t$ given $B_{j}$ at time $0$ ($i,j \! = \! 1,2,3,4$) \end{center}
for the left going ($l$) meson, and

\begin{center} $q_{ij}(t|0) =$ probability of $B_{i}$ at time $t$ given $B_{j}$ at time $0$ ($i,j \! = \! 1,2,3,4$) \end{center}
for the right going ($r$) meson. The symbols $p$ and $q$ will be used for all the probabilities of the left and right going meson respectively.

It can be shown that all the physical conditions imposed by QM to single $B$-mesons are satisfied by the probabilities of LR which can be collected in a ``probability matrix'' 
\begin{eqnarray} M= \left( \begin{array} {cccc} E(t)Q_{+}(t) & E(t)Q_{-}(t) & 0 & 0 \\ E(t)Q_{-}(t) & E(t)Q_{+}(t) & 0 & 0 \\ 0 & 0 & E(t)Q_{+}(t) & E(t)Q_{-}(t) \\ 0 & 0 & E(t)Q_{-}(t) & E(t)Q_{+}(t) \end{array} \right) \end{eqnarray} 
where $p_{11}(t|0) = q_{11}(t|0) = E(t)Q_{+}(t)$, $p_{12}(t|0) = q_{12}(t|0) = E(t)Q_{-}(t)$, etc., and

\begin{equation} Q_{\pm}(t) \equiv \frac{1}{2} [1 \pm cos(\Delta m t)] \end{equation}
so that $Q_{+}(t) \! + \! Q_{-}(t) \! = \! 1$. The proof of (13) is strictly analogous to that for neutral kaons published in [9]. It has been shown that this set of probabilities is unique within the local realistic approach [6].

\subsection{Three time probabilities}
\indent

The physical situation described in QM by the wave function (1) corresponds to the initial LR probabilities

\begin{equation} q_{1}(0) = q_{2}(0) = q_{3}(0) = q_{4}(0) =  \frac{1}{4} \end{equation}
where $q_{i}$ is the probability of the state $B_{i}$ ($i \! = \! 1,2,3,4$). Our present task is to find the maximum and minimum values allowed by LR for the asymmetry parameter $A$, for $t_{r} \! > \! t_{l}$. Two time probabilities can be written by means of probabilities with three indices by using Bayes' formula [16]:

\begin{eqnarray} \begin{array}{ll} q_{11}(t_{r},t_{l}) = q_{111}(t_{r},t_{l}|0)q_{1}(0) + q_{112}(t_{r},t_{l}|0)q_{2}(0) \\ q_{21}(t_{r},t_{l}) = q_{211}(t_{r},t_{l}|0)q_{1}(0) + q_{212}(t_{r},t_{l}|0)q_{2}(0) \end{array} \end{eqnarray}
where $q_{ijk}(t_{r},t_{l}|0)$ is the probability of having $B_{i}$ at time $t_{r}$ and $B_{j}$ at time $t_{l}$ given a $B_{k}$ at time $0$ ($i,j,k=1,2,3,4$). To simplify the notation, let us set $E_{l} \! \equiv \! E(t_{l})$, $Q_{\pm}^{l} \! \equiv \! Q_{\pm}(t_{l})$ and  $E_{r} \! \equiv \! E(t_{r})$, $Q_{\pm}^{r} \! \equiv \! Q_{\pm}(t_{r})$. One must have

\begin{equation} q_{111}(t_{r},t_{l}|0) + q_{121}(t_{r},t_{l}|0) = E_{r}Q_{+}^{r} \end{equation} 
\begin{equation} q_{211}(t_{r},t_{l}|0) + q_{221}(t_{r},t_{l}|0) = E_{r}Q_{-}^{r} \end{equation}
because the left hand sides are the probabilities of $B_{i}$ at time $t_{r}$ ($i \! = \! 1,2$) summed over all possible states at time $t_{l}$, given a $B_{1}$ at time 0. Therefore the left hand sides of (17) and (18) must equal $q_{11}(t_{r}|0)$  and $q_{21}(t_{r}|0)$, which are given by (13). To (17) and (18) we can add

\begin{equation} q_{111}(t_{r},t_{l}|0) + q_{211}(t_{r},t_{l}|0) = E_{r-l}q_{11}(t_{l}|0) \end{equation}
\begin{equation} q_{121}(t_{r},t_{l}|0) + q_{221}(t_{r},t_{l}|0) = E_{r-l}q_{21}(t_{l}|0) \end{equation}
where $E_{r-l} \! \equiv \! E(t_{r} \! - \! t_{l})$. The left hand side of (19) is the probability that the meson is either $B_{1}(t_{r})$ or $B_{2}(t_{r})$ (then that is undecayed at $t_{r}$) and that it is $B_{1}(t_{l})$, given that it was $B_{1}(0)$. The meaning of (20) is similar. The probabilities on the right hand sides of (19) and (20) are again given by (13). Therefore (19) and (20) can be written as

\begin{equation} q_{111}(t_{r},t_{l}|0) + q_{211}(t_{r},t_{l}|0) = E_{r}Q_{+}^{l} \end{equation} 
\begin{equation} q_{121}(t_{r},t_{l}|0) + q_{221}(t_{r},t_{l}|0) = E_{r}Q_{-}^{l} \end{equation}
because of $E_{r} \! = \! E_{r-l}E_{l}$. Four unknown probabilities appear in (17), (18) and (21), (22). These equations are however not independent because the sum of (17) and (18) equals the sum of (21) and (22) (remember that $Q_{+} \! + \! Q_{-} \! = \! 1$). One can write three probabilities in terms of $q_{111}$:

\begin{eqnarray} \left\{ \begin{array}{lll} q_{121} = E_{r}Q_{+}^{r}-q_{111}   \\  q_{211} = E_{r}Q_{+}^{l}-q_{111} \\  q_{221} = E_{r}(Q_{-}^{l}-Q_{+}^{r})+q_{111} \end{array} \right.  \end{eqnarray}
where the arguments $(t_{r},t_{l}|0)$ of the three time probabilities have been omitted, as it will be done systematically in the following. Obvious upper and lower limits for $q_{111}$ are

\begin{equation} 0 \leq q_{111} \leq E_{r}Q_{+}^{r}, E_{r}Q_{+}^{l} \end{equation}
as follow from (17) and (21). Analogous limits can be obtained for the other three probabilities which, using (23), can be translated into further limitations for $q_{111}$. The overall result is

\begin{equation} p \leq q_{111} \leq P \end{equation}
where
\begin{equation} p = E_{r} Max\{0;(Q_{+}^{r}-Q_{-}^{l})\} \hspace{0.5cm} , \hspace{0.5cm}   P = E_{r} Min\{Q_{+}^{l};Q_{+}^{r}\} \end{equation}
When different limitations are given it is useful to adopt the most stringent ones. For this reason in (26) the maximum of the minima and the minimum of the maxima are considered. Which one of the two terms within curly brackets has to be chosen depends on $t_{l}$ and $t_{r}$; for instance $Q_{+}^{r} \! - \! Q_{-}^{l}$ can be positive or negative depending on the considered times. Equations similar to (17), (18) and (21), (22) can be written for an initial $B_{2}$. They are:

\begin{equation} q_{222} + q_{212} = E_{r}Q_{+}^{r} \end{equation}
\begin{equation} q_{122} + q_{112} = E_{r}Q_{-}^{r} \end{equation} 
and
\begin{equation} q_{222}+ q_{122} = E_{r}Q_{+}^{l} \end{equation} 
\begin{equation} q_{212} + q_{112} = E_{r}Q_{-}^{l} \end{equation} 
Notice that if the indeces 1 and 2 are systematically interchanged (17), (18) and (21), (22) become (27), (28) and (29), (30) respectively. Conclusions similar to (23)-(25) must then hold for these new probabilities, by applying the same interchange. One has

\begin{eqnarray} \left\{ \begin{array}{lll} q_{212} = E_{r}Q_{+}^{r}-q_{222}   \\  q_{122} = E_{r}Q_{+}^{l}-q_{222} \\  q_{112} = E_{r}(Q_{-}^{l}-Q_{+}^{r})+q_{222} \end{array} \right.  \end{eqnarray}
and

\begin{equation} p \leq q_{222} \leq P \end{equation}
with $p$ and $P$ again given by (26). Remembering (16) and (15), and using (23) and (31), one gets

\begin{equation} q_{11}(t_{r},t_{l}) = \frac{1}{4}(q_{111}+ q_{112}) = \frac{1}{4} [E_{r}(Q_{-}^{l}-Q_{+}^{r})+q_{111}+q_{222}] \end{equation}
\begin{equation} q_{21}(t_{r},t_{l}) = \frac{1}{4}(q_{211}+ q_{212}) = \frac{1}{4} [E_{r}(Q_{+}^{l}+Q_{+}^{r})-q_{111}-q_{222}] \end{equation}
Similar reasonings can be made for $B_{3}(0)$ and $B_{4}(0)$. The results are very similar to the previous case and one obtains

\begin{equation} q_{33}(t_{r},t_{l}) = \frac{1}{4}(q_{333}+ q_{334}) = \frac{1}{4} [E_{r}(Q_{-}^{l}-Q_{+}^{r})+q_{333}+q_{444}] \end{equation}
\begin{equation} q_{43}(t_{r},t_{l}) = \frac{1}{4}(q_{433}+ q_{434}) = \frac{1}{4} [E_{r}(Q_{+}^{l}+Q_{+}^{r})-q_{333}-q_{444}] \end{equation}
with the limitations

\begin{equation} p \leq q_{333} \leq P \hspace{0.5cm} , \hspace{0.5cm} p \leq q_{444} \leq P \end{equation}
with $p$ and $P$ again given by (26).

\subsection{B-meson pair probabilities}
\indent

Next we calculate the local realistic probability of having two $B^{0}$ mesons with $b \! = \! -1$ at proper times $t_{l}$ and $t_{r} \! > \! t_{l}$. We must then consider the situations in which there is either $B_{2}$ or $B_{4}$ on the left at time $t_{l}$ and on the right at time $t_{r}$. This probability is:

\begin{equation} P^{LR}[\overline{B^{0}}(t_{l});\overline{B^{0}}(t_{r})] = E_{l}q_{21}(t_{r}, t_{l}) + E_{l}q_{43}(t_{r}, t_{l}) \end{equation}
The first term in (38) is the probability that the first $B^{0}$ is undecayed at time $t_{l}$ times the probability that the second $B^{0}$ is $B_{2}(t_{r})$ and $B_{1}(t_{l})$: the latter factor, due to anti-correlation, coincides with the probability that the second $B^{0}$ is $B_{2}(t_{r})$ and the first one  $B_{4}(t_{l})$. Similar is the meaning of the second term in (38). We must also calculate the probability of having a $B^{0}$ with $b \! = \! +1$ at proper time $t_{l}$ and a $B^{0}$ with $b \! = \! -1$ at proper time $t_{r} \! > \! t_{l}$. It is clearly given by

\begin{equation} P^{LR}[\overline{B^{0}}(t_{l});B^{0}(t_{r})] = E_{l}q_{33}(t_{r}, t_{l}) + E_{l}q_{11}(t_{r},t_{l}) \end{equation}
From (38) and (39) it follows

\begin{equation} P^{LR}[\overline{B^{0}}(t_{l});B^{0}(t_{r})] + P^{LR}[\overline{B^{0}}(t_{l});\overline{B^{0}}(t_{r})] = E_{l} [q_{21}+q_{11}+q_{43}+q_{33}] \end{equation}

From (33)-(36) one also has
\begin{equation} P^{LR}[\overline{B^{0}}(t_{l});B^{0}(t_{r})] + P^{LR}[\overline{B^{0}}(t_{l});\overline{B^{0}}(t_{r})] = \frac{1}{2}E_{r+l} \end{equation}
Furthermore

\begin{equation} P^{LR}[\overline{B^{0}}(t_{l});B^{0}(t_{r})] - P^{LR}[\overline{B^{0}}(t_{l});\overline{B^{0}}(t_{r})] = E_{l}[q_{33}-q_{43}]+ E_{l}[q_{11}-q_{21}] \end{equation}

Using (33)-(36) and the limits (25), (32) and (37) it follows

\begin{equation} \{P^{LR}[\overline{B^{0}}(t_{l});B^{0}(t_{r})] - P^{LR}[\overline{B^{0}}(t_{l});\overline{B^{0}}(t_{r})] \}_{max} = \frac{1}{2} E_{r+l}(1-2\mid Q^{l}_{+} - Q^{r}_{+} \mid) \end{equation}
and

\begin{equation} \{\! P^{LR}[\overline{B^{0}}(t_{l});B^{0}(t_{r})] - P^{LR}[\overline{B^{0}}(t_{l});\overline{B^{0}}(t_{r})] \}_{max} \\ = \frac{1}{2} E_{r+l} \{1 \! - \! 2Min[(Q^{l}_{+} + Q^{r}_{+});(Q^{l}_{-} - Q^{r}_{-})]\} \end{equation}
Thus, for LR, the asymmetry parameter (9) turns out to have maximum and minimum values given by

\begin{eqnarray} \left\{ \begin{array}{ll} A^{LR}_{max}(t_{l};t_{r}) = 1-2\mid Q^{l}_{+} - Q^{r}_{+} \mid \\ A^{LR}_{min}(t_{l};t_{r}) = 1-2Min[(Q^{l}_{+} + Q^{r}_{+});(Q^{l}_{-} + Q^{r}_{-})] \end{array} \right. \end{eqnarray}

Eq. (45) can be usefully written again with the explicit dependence on $(t_{r} \! - \! t_{l})/\tau \! \equiv \! \eta$ (with $\eta \! > \! 0$):

\begin{eqnarray} \left\{ \begin{array}{ll} A^{LR}_{max}(t_{l};t_{r}) = 1 - \mid (1-cos(x\eta))cos(xt_{l}) + sin(x\eta)sin(xt_{l}) \mid \\ A^{LR}_{min}(t_{l};t_{r}) = 1 - Min [2+\Psi;2-\Psi] \end{array} \right. \end{eqnarray}
where

\begin{equation} \Psi=(1+cos(x\eta))cos(xt_{l})-sin(x\eta)sin(xt_{l}) \end{equation}

The behaviour of (10) and (45) is shown in Fig. 1. 

It can be noticed that $A^{QM}$ depends only on the difference of the two proper times and remains unchanged under the $t_{l} \! \leftrightarrow \! t_{r}$ exchange (thus showing a symmetric behaviour with respect to $(t_{r} \! - \! t_{l})/\tau$). Therefore with $t_{l} \! > \! t_{r}$, the study of $A^{QM}$ is reduced to the case $t_{r} \! > \! t_{l}$. On the other hand  $A^{LR}$ depends not only on this difference but also on one given absolute proper time, $t^{\ast}$. However, under the $t_{l} \! \leftrightarrow \! t_{r}$ exchange $A^{LR}$ is symmetric as can be easily shown and as should be expected from the conventionality of $l$ and $r$ tags. Thus the study of $A^{LR}$ can be limited to the case $t_{r} \! \geq \! t_{l}$ establishing, conventionally, the dependence of $A^{LR}$ on the shorter absolute time. 

To appreciate the difference between the prediction of QM and the maximum prediction of LR it is useful to represent the asymmetry parameter as a function of $(t_{r} \! - \! t_{l})/ \tau$ for some given values of $t_{l}$ as in Fig. 2. 

The comparison of the whole LR prediction with the MQ one, for two representative values of $t_{l}$, is given in Fig. 3.

It is interesting to notice that the difference between the predictions of QM and LR is higher in the $B^{0} \! - \! \overline{B^{0}}$ system than in the $K^{0} \! - \! \overline{K^{0}}$ system, after comparison with the results of ref. [10]. However from an experimental point of view B-mesons require measuring smaller decay times that is more difficult.

\section{EXPERIMENTAL FEASIBILITY OF THE EPR TEST}
\indent

The construction of the asymmetric B-factories (high luminosity asymmetric $e^{+}e^{-}$ colliders operating at the $\Upsilon(4S)$ resonance), i.e. the PEP-II storage ring instrumented with \textit{BaBar} detector at SLAC [17], and the KEK-B storage ring with \textit{Belle} detector at KEK [18], will provide a powerful tool to perform the EPR test just proposed. Indeed, as we are going to argue, this test can be performed with the experimental setups and software tools already being developed for CP-violation, lifetime and mixing measurements, and the precision required for these measurements is adequate. Firstly we review some characteristics and experimental parameters concerning asymmetric B-factories, then we suggest two methods to perform the EPR test and finally we provide a quantitative evaluation for the test feasibility.

\subsection{Methods and performances at the asymmetric B-factories}
\indent

The asymmetry (9), as predicted by QM, depends on the difference of the two B-meson proper decay times $\Delta t$. The asymmetric collider configuration and an accurate vertex determination make the $\Delta t$ determination possible. Indeed, the fundamental reason for developing asymmetric B-factories is the need to measure small proper time intervals, whereas a precise vertex detector is necessary to measure a space dependence from which the time dependence can be deduced. Time measurements would not be feasible if the $\Upsilon(4S)$, having energy just over the $B_{d}$ meson pair production threshold, were produced at rest: the B mesons have very low momentum ($340MeV \! / c$) in the center-of-mass frame and the paths between production and decays are too small ($30 \mu m$ in average) to be measured even by silicon vertex detectors. Therefore proper decay time measurements require the asymmetric beam configuration that produces the $\Upsilon(4S)$ resonance in motion in the laboratory frame, thus allowing the expansion between the two decay positions. Neglecting B-mesons motion in the $\Upsilon(4S)$-frame, the proper time difference is in good approximation given by $\Delta t \! \simeq \! \Delta z / c \eta$ where $\Delta z$ is the distance between the two B decay vertices measured in the laboratory, along the beams direction, and $\eta \! = \! \beta \gamma$ is the Lorentz boost factor due to the asymmetric beam energies. Its value is chosen as a good compromise between the $\Delta z$ expansion and the necessity that the boosted decay products are not lost in the dead cone ahead avoiding efficiencies decrease: $\eta \! = \! 0.56$ at \textit{BaBar}-PEP II [17], $\eta \! = \! 0.42$ at \textit{Belle}-KEKB [18]. The applied boost expands average separation of the two B-vertices $\Delta z$ to about $250 \mu m$. In this case a silicon vertex detector allows to determine the two B-decay vertices separation $\Delta z$. Both \textit{BaBar} [17] and \textit{Belle} [18] can provide $\sigma_{\Delta z}$ of the order of $100 \mu m$ and in some case better, depending on the detected decay channels.

Among others three B-physics measurements can be performed at asymmetric B-factories: CP-asymmetries, $B_{d}$ mixing and $B_{d}$ lifetime. Both the physical time-dependent asymmetries, that need to be measured by \textit{BaBar} and \textit{Belle} in order to establish CP violation and determine the mixing parameter $\Delta m_{B}$, depend on the difference $\Delta t$ between the two B-meson proper decay times. 

To measure CP asymmetries one looks for events where one neutral B meson decays into an hadronic CP-eigenstate at the time $t_{CP}$ while the other decays, at $t_{tag}$, to a semi-leptonic or hadronic \textit{tagging mode} that acts as a b-flavour identifier:  $\Delta t \! = \! t_{CP}\! - \! t_{tag}$. 

On the other hand mixing measurement can be based on double tagged dileptonic events. It must be pointed out that the dilepton approach with a time-dependent asymmetry is different from the usual time-integrated dilepton methods developed at a symmetric $e^{+}e^{-}$ collider at $\Upsilon(4S)$ (by CLEO experiment) which allows only the measurement of $\chi_{d}=x^{2}_{d}/2(1+x^{2}_{d})$.

Most lifetime measurement methods rely on the determination of both the secondary vertices and of the primary one, the $\Upsilon(4S)$ decay point; an exclusive reconstruction for at least one B meson is needed, while the other B can be partially reconstructed or only vertexed.

In general the resolution achievable in individual tracks reconstruction is limited by multiple scattering introduced primarily by the beam pipe. The resolution $\sigma_{\Delta z}$ (for example $\sigma_{\Delta z} \! = \! \sqrt{\sigma_{z_{CP}}^{2} \! + \! \sigma_{z_{tag}}^{2}}$) is primarily determined by the error in tagged vertex position $\sigma_{z_{tag}}$ [17,18]. 

Typical values are $\sigma_{z} \! \simeq \! 50 \mu m$ for a fully reconstructed B meson and $\sigma_{z} \! \simeq \! 100 \mu m$ for a partially reconstructed one. In clean hadronic channels and in lepton tagging better resolution can be achieved in principle thus together providing $\sigma_{\Delta z}$ better than $100 \mu m$ for the CP analysis and $\sigma_{\Delta z} \! \simeq \! 110 \mu m$ in the dilepton analysis. On the other hand non-gaussian tails with $\sigma \! \sim \! 200 \! \div \! 300 \mu m$ for a significant fraction of the selected samples may partially spoil these resolutions.

In conclusion we assume that the asymmetric machine configuration and high performance in B vertexing will provide time-dependent measurements with $\sigma_{\Delta z}$ of the order of $100 \! \div \! 110 \mu m$ so that $\sigma_{\Delta t}\! = \! \sigma_{\Delta z}/ \beta \gamma c \sim 0.4 \, \tau$.
 
\subsection{EPR test performed by lifetime measurement method}
\indent

The $\Upsilon(4S)$ decays almost exclusively to pairs of B mesons, that are neutral with a probability of about $50\%$. In such cases no ionizing particle comes out of the $\Upsilon (4S)$ decay point which is thus not directly measurable. This in turn does not permit to record and use decay distances in lifetime determination. However lifetime measurements can be performed, without knowing the B-mesons production point, using only the $\Delta z$ measurement. On the other hand a more sophisticated method of reconstructing the geometry and the kinematics of the events has been developed in \textit{BaBar} in order to perform more direct lifetime measurements [17]. This method can suggest how to perform a measurement of the asymmetry (9) by measuring not only $\Delta t$ but also one time of flight ($t^{*} \! = \! t_{l}$, subsection $3.3$) as needed by the local realistic time dependence. 

In this method the decay vertices of the fully reconstructed $B^0$, of the other $B^0$ and the $\Upsilon(4S)$ production point, considered to belong the very flat ribbon-shaped beam spot ($\sigma_{x} \! \simeq \! 140 \mu m$, $\sigma_{y} \! \simeq \! 6 \mu m$, $\sigma_{z} \! \simeq \! 1 cm$), are connected assuming the direction of their lines of flight from kinematics. Thus a gain in measurement accuracy is provided by adding kinematical constraints to the standard geometrical fit of the decay vertices. 

To perform the EPR test the quantities $\sigma_{\Delta t}$, $\sigma_{t^{*}}$ must be both small enough in order to allow an effective time-dependent analysis. Qualitatively $\sigma_{\Delta z}$ and $\sigma_{z-z_{\Upsilon}}$ should be smaller than half the average separation between vertices, and the typical value of $\sigma_{\Delta z} \! \sim \! 100 \mu m$ ($\sigma_{\Delta t} \! \sim \! \! 0.4 \, \tau$) could be adequate. This requirement could be satisfied with both the B mesons fully reconstructed and using the algorithm mentioned above. On the other hand, as the rate of fully reconstructible B mesons is likely to be of the order of $1\%$ or less [17], the second decay should not be fully reconstructed in order to get sufficient statistics in a relatively short time. However for this second decay no vertexing technique can provide a $\sigma_{z-z_{\Upsilon}}$ smaller than $100 \mu m$. From the point of view of the EPR test it is preferable to use the $t^{*} \! = \! t_{l}$ time only if it belongs to the fully reconstructed B meson. This is only satisfied on average in one half of the events and reduces the statistics by one half but to a level still higher than achievable considering both fully reconstructed B mesons.

\subsection{EPR test performed by mixing measurement method}
\indent

A way to experimentally obtain the asymmetry (9), in order to test the quantum mechanical and local realistic predictions, (10) and (45) respectively, can be based on the measurement of the time-dependent mixing.

Since $P[B^{0}(t_{l}); \overline{B^{0}}(t_{r})]=P[\overline{B^{0}}(t_{l}); B^{0}(t_{r})], P[\overline{B^{0}}(t_{l}); \overline{B^{0}}(t_{r})]=P[B^{0}(t_{l}); B^{0}(t_{r})]$ hold both in QM and in LR, the asymmetry (9) can be written as 

\begin{equation} A(t_{l},t_{r}) = \frac{ P[B^{0}(t_{l}); \overline{B^{0}}(t_{r})]+P[\overline{B^{0}}(t_{l}); B^{0}(t_{r})]-P[\overline{B^{0}}(t_{l}); \overline{B^{0}}(t_{r})]-P[B^{0}(t_{l}); B^{0}(t_{r})] } { P[B^{0}(t_{l});\overline{B^{0}}(t_{r})]+P[\overline{B^{0}}(t_{l}); B^{0}(t_{r})]+P[\overline{B^{0}}(t_{l});\overline{B^{0}}(t_{r})]+P[B^{0}(t_{l}); B^{0}(t_{r})] }\end{equation}
and also as

\begin{equation} A(t_{l},t_{r}) = \frac {N(B^{0},\overline{B^{0}})(t_{l},t_{r})-N(\overline{B^{0}},\overline{B^{0}})(t_{l},t_{r})-N(B^{0},B^{0})(t_{l},t_{r}) } {N(B^{0},\overline{B^{0}})(t_{l},t_{r})+N(\overline{B^{0}},\overline{B^{0}})(t_{l},t_{r})+N(B^{0},B^{0})(t_{l},t_{r}) } \end{equation}
where $N(B^{0},\overline{B^{0}})$ represents the number of unlike-flavour events.

In LR the asymmetry depends not only on $\Delta t$, as in QM, but also on $t^{\ast}$. However the local realistic time dependence on  $\Delta t$ could be directly compared to that of QM, for instance by performing a partial time integration, namely an integration on $t_{l}/\tau$ for $t_{l}/\tau \! < \! 2$ with the constraint $\Delta t/\tau \! \leq \! 2$. Moreover, considering that a very large fraction of the events (double B-decays) is included in the double requirement ($t_{l}/\tau \! < \! 3$ and $t_{r}/\tau \! < \! 3$), for a given $t_{l}/\tau \! < \! 3$ these events are characterized by values of $\Delta t$ belonging to the time interval $[\, 0, \, 3 \! - \! (t_{l}/\tau)]$, for which the local realistic prediction is always below the quantum mechanical one. Thus a total integration on $t_{l}/\tau$ is clearly allowed. This permits to consider the following quantity simplified in its time dependence:

\begin{equation} A(\Delta t) = \frac {N(B^{0},\overline{B^{0}})(\Delta t)-N(\overline{B^{0}},\overline{B^{0}})(\Delta t)-N(B^{0},B^{0})(\Delta t) } {N(B^{0},\overline{B^{0}})(\Delta t)+N(\overline{B^{0}},\overline{B^{0}})(\Delta t)+N(B^{0},B^{0})(\Delta t) } \end{equation}

Therefore, considering a dilepton tagging approach to obtain experimentally the asymmetry (50), one can measure the asymmetry 

\begin{equation} A(\mid \! \Delta t \! \mid) = \frac {N(l^{+}l^{-})-N(l^{+}l^{+})-N(l^{-}l^{-})} {N(l^{+}l^{-})+N(l^{+}l^{+})+N(l^{-}l^{-})} \end{equation} 
by counting the number of \textit{like-sign} and \textit{unlike-sign} events, $N(l^{\pm}l^{\pm})$ and $N(l^{+}l^{-}$), as a function of $\mid \! \Delta t \! \mid$.

As the $\Delta z$ resolution for the cascade leptons is too large, in the mixing study the B-meson flavour is correlated only with the sign of the direct leptons (distinguished by their harder momentum spectrum), and the cascade ones act as a source of \textit{mistagging} [17]. To obtain the yield of dilepton events from semi-leptonic decays of $B^{0} \overline{B^{0}}$ pairs, several other backgrounds must be subtracted, such as dileptons from the non resonant process at $\Upsilon(4S)$ (\textit{continuum}) and hadrons misidentified as leptons (\textit{fake leptons}).

\subsection{Quantitative evaluation for test feasibility}
\indent

Two methods, based on lifetime and mixing measurements, have been proposed to perform the EPR test. The former, needing time-dependent measurement not only for $\Delta t$ but also for $t^{*}$, requires an exclusive channels approach. Its need of larger statistics suggests the choice of the latter method to experimentally obtain the asymmetry (9) for an initial study. 

The \textit{BaBar} relative accuracy of about $1\%$ in $\Delta m_{B}$ [17] with only one year data taking at nominal luminosity ($30fb^{-1}$) suggests that at the same time an accuracy on the asymmetry (9) largely sufficient to discriminate between QM and LR predictions should be obtained. 

The number of events required to obtain a given level of separation between the two theoretical predictions provides a quantitative evaluation of the power of the proposed test. The number of $B^{0} \overline{B^{0}}$ events to be produced, in order to measure an asymmetry $A$ with a statistical error $\delta A$ is approximately given by

\begin{equation} N_{prod} \approx \frac{1}{(\delta A)^2} \cdot \frac{1}{D^2 \epsilon Br} \end{equation}
where $D$ represents the product of all dilution factors, $\epsilon$ is the total detection efficiency and $Br$ ($\simeq \! 0.04$) is the branching ratio into the dileptonic final state.
The corresponding needed luminosity integrated over one typical year is

\begin{equation} \displaystyle{\mathcal{L}} \equiv \int L dt = \frac{N_{prod}}{\sigma_{b \overline{b}}} \end{equation}
where $\sigma_{b \overline{b}} \! \simeq \! 1.2 \,nb$ is the cross section for $b \overline{b}$ production at the $\Upsilon(4S)$.
Neglecting higher order effects due to backgrounds, the total dilution factor $D$ is simply given by the tagging dilution factor $d_{tag} \! = \! (1 \! - \! 2\eta)$ where the \textit{mistag probability} $\eta$ is the probability that a $B^{0}\overline{B^{0}}$ pair is tagged as a $B^{0}B^{0}$ or $\overline{B^{0}}\overline{B^{0}}$ pair. Typical values are $\eta \! = \! 13\%$ and $\epsilon \! = \! 45\%$ [17].

$N_{prod}$ has been evaluated requiring the separation between the integrated values of $A_{QM}$ and $A_{LR}$ to be at least $4\sigma$ over an appropriate interval of the decay time distribution for $B^{0}\overline{B^{0}}$ pairs. Considering that $\Delta t$ and $t^{\ast}$ both greater than $0.3 \, \tau$ give a large difference between the two predictions, one obtains $\textstyle{\mathcal{L}} \approx 1fb^{-1}$. This is enough to discriminate between the two theoretical predictions. The larger luminosity of $30fb^{-1}$, integrated in one nominal year, will allow a more detailed and stringent test.

\section{CONCLUSIONS}
\indent

Developing a very general local realistic theory of correlated $B^{0}_{d}\overline{B^{0}_{d}}$ a not-of-Bell-type EPR test has been proposed to discriminate between quantum mechanics and local realism. Indeed the asymmetry for observing neutral $B_{d}$ meson pairs with like and unlike flavour predicted by local realism is remarkably different from the quantum mechanical prediction.

This test can be based on mixing and lifetime measurements to be carried out by the experiments at the asymmetric B-factories and therefore is not only experimentally feasible but can be effectively performed quite soon if the expectations on the colliders luminosity and the detectors performance will be fulfilled by \textit{BaBar} and \textit{Belle} collaborations.

\section{ACKNOWLEDGEMENTS}
\indent

We would like to thank L. Lanceri for fruitful discussions on the experimental test feasibility and G. Nardulli for his comments.

\section{REFERENCES}
\noindent
\begin{enumerate}
\item A. Einstein, B. Podolsky, and N. Rosen, Phys. Rev. \textbf{A 47}, 777 (1935).
\vspace{-0.3cm}
\item J.S.Bell, Physics \textbf{1}, 195 (1965).
\vspace{-0.3cm}
\item J.F. Clauser, M.A. Horne, A. Shimony, and R.A. Holt, Phys. Rev. Lett. \textbf{23}, 880 (1969).
\vspace{-0.3cm} \item S.J. Freedman and J.F. Clauser, Phys. Rev. Lett. \textbf{28}, 938 (1972);
E.S. Fry and R.C. Thompson, Phys. Rev. Lett. \textbf{37}, 465 (1976); A. Aspect, P.Grangier and G. Roger, Phys. Rev. Lett. \textbf{47}, 460 (1981); \textbf{49}, 91 (1982); \textbf{49}, 1804 (1982); W. Perrie, A.J. Duncan, H.J. Beyer and H. Kleinpoppen, Phys. Rev. Lett. \textbf{54}, 1790 (1982).
\vspace{-0.3cm} \item V.L. Lepore and F. Selleri, Found. Phys. Lett. \textbf{3}, 203 (1990).
\vspace{-0.3cm} \item A. Afriat and F. Selleri, \textit{The Einstein, Podolsky and Rosen paradox in atomic, nuclear and particle physics}, Plenum, London and New York, 1998.
\vspace{-0.3cm} \item G.C. Ghirardi, R. Grassi and T. Weber, in \textit{The DA$\Phi$NE Physics Handbook}, vol \textbf{I}, edited by L. Maiani, G.Pancheri and N. Paver, INFN, Frascati, 1992.
\vspace{-0.3cm} \item Most of them are relative to $K^{0}\overline{K^{0}}$ pairs and $\Phi$-factories. Among them:  P.H. Eberhard, Nucl. Phys. \textbf{B 398}, 155 (1993); A. Di Domenico, Nucl. Phys. \textbf{B 450}, 293 (1995); B. Ancochea, A. Bramon and M. Nowakowski, hep-ph/9811404; F. Benatti and R. Floreanini, Phys. Rev. \textbf{D 57}, 1332 (1998). With regard to $B^{0}\overline{B^{0}}$ pairs: A. Datta, in \textit{Proceedings of the Workshop on High Energy Physics Phenomenology}, Calcutta, 1996.
\vspace{-0.3cm} \item F. Selleri, Phys. Rev. \textbf{A 56}, 3493 (1997).
\vspace{-0.3cm} \item R. Foad\'{i} and F. Selleri, submitted to Phys. Letters.
\vspace{-0.3cm} \item R.A. Bertlmann and W. Grimus, Phys. Lett. \textbf{B 392}, 426 (1997); G.V. Dass and K.V.L. Sarma, Eur. Phys. J. \textbf{C5 (2)}, 283 (1998); R. Bertlmann and W. Grimus, Phys. Rev. \textbf{D 58}, 034014 (1998). 
\vspace{-0.3cm} \item D. Bohm and Y. Aharonov, Phys. Rev. \textbf{108}, 1070 (1957).
\vspace{-0.3cm} \item Strong inequalities that are necessary consequences of the SFH were falsified in the experiments of ref. [4]. 
\vspace{-0.3cm} \item CPLEAR Collaboration (A. Apostolakis et al.), Phys. Lett. \textbf{B 422}, 339 (1998). 
\vspace{-0.3cm} \item C. Caso \textit{et al.} \textit{Review of Particle Physics}, Eur. Phys. J. \textbf{C3}, 1 (1998). 
\vspace{-0.3cm} \item R. Foad\'{i}, Master thesis, Universit\`{a} di Torino, 1998. 
\vspace{-0.3cm} \item The BaBar Collaboration, \textit{BaBar Physics Book}, \textbf{SLAC-R-504}, 1998. 
\vspace{-0.3cm} \item BELLE Collaboration, \textit{Technical Design Report}, \textbf{KEK-R-95-1}, 1995. 
\end{enumerate}

\clearpage

\begin{figure}[ht]
\begin{tabular}{cc} 
\scalebox{.38}{\includegraphics{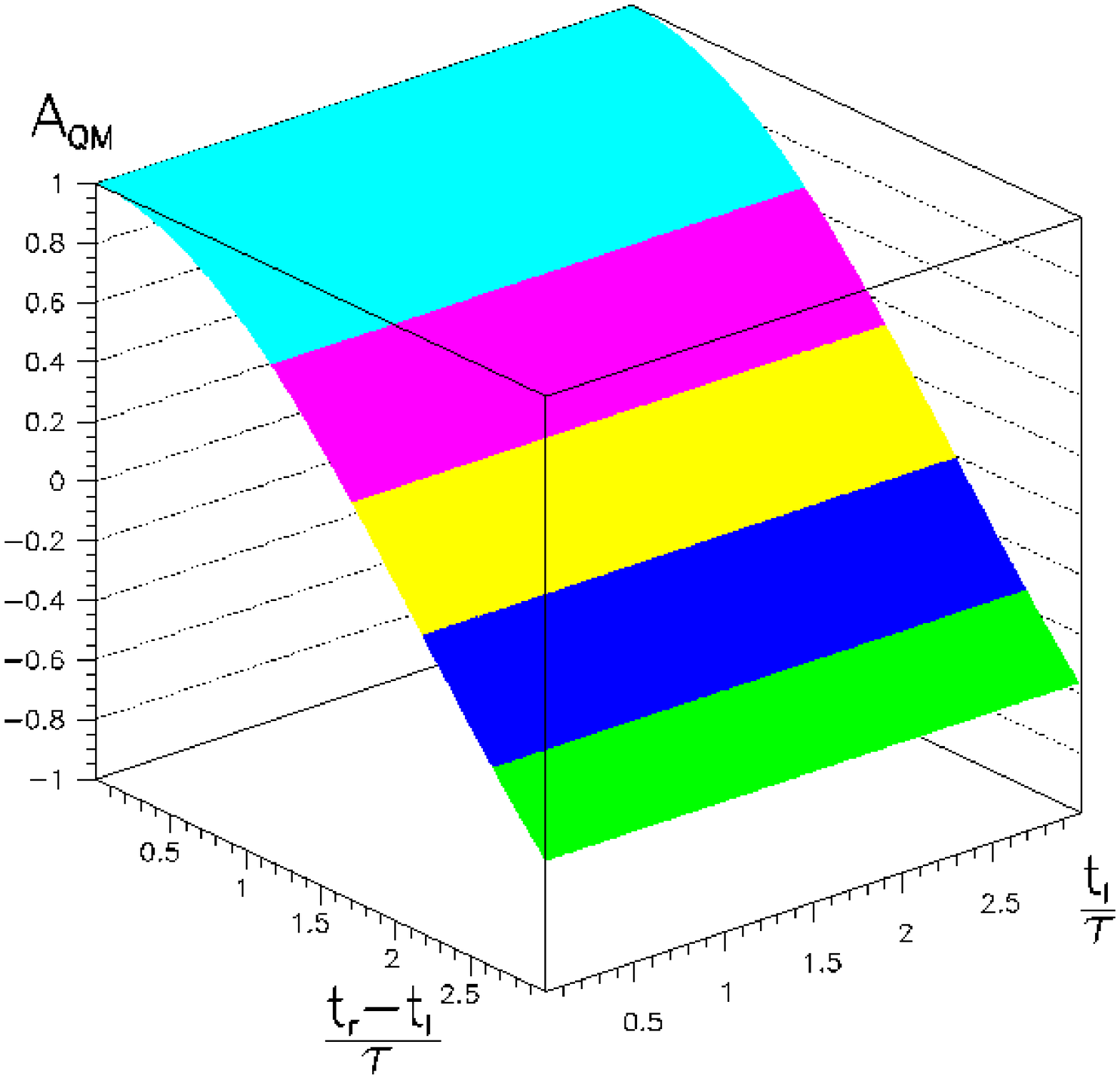}} &
\scalebox{.38}{\includegraphics{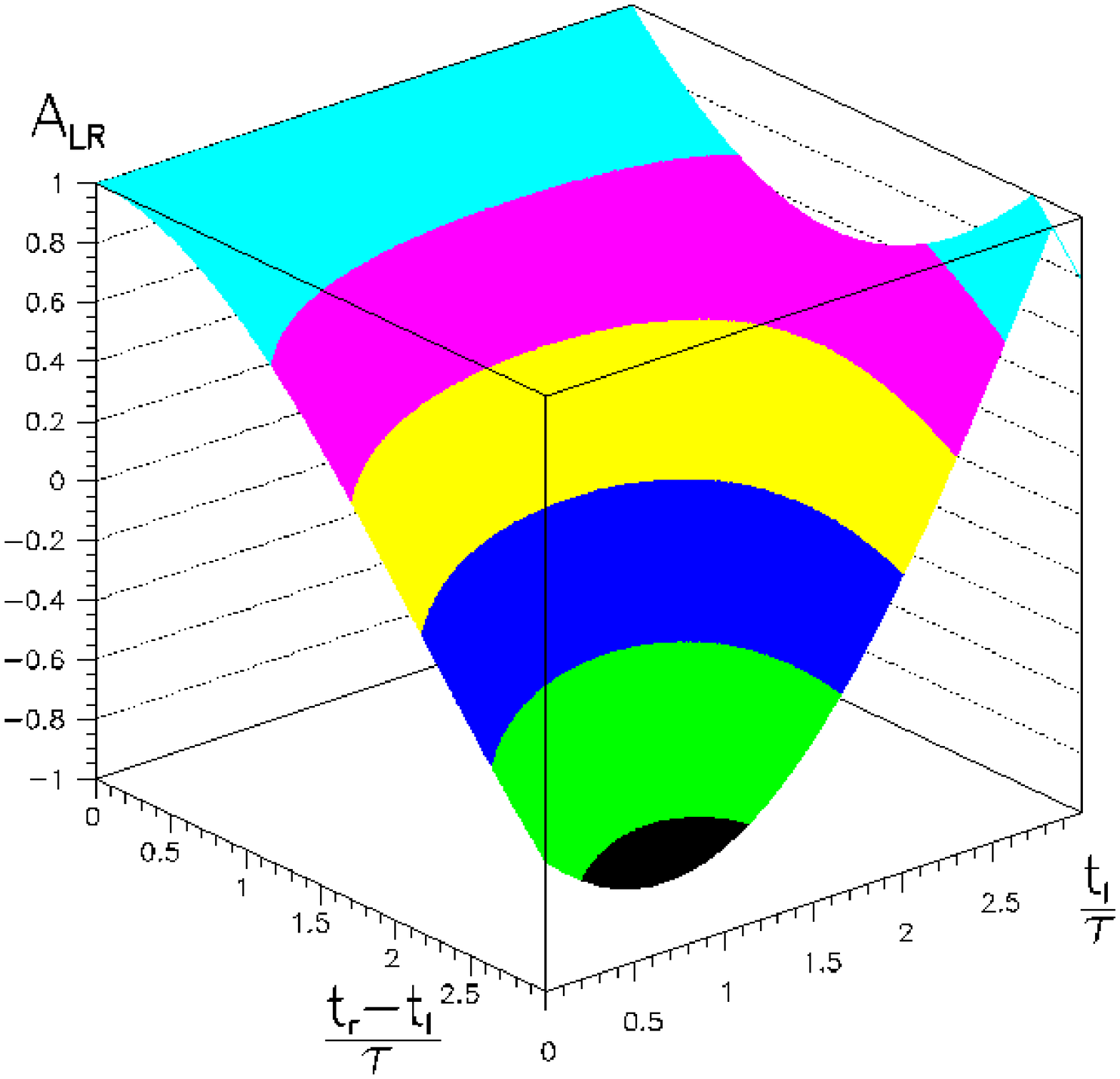}} \\
\end{tabular}
\caption{\label{fig1}Flavour asymmetry predicted by QM (left) and LR (right; maximum values).}
\end{figure}

\clearpage

\begin{figure}[ht]
\begin{center}
\scalebox{.4}{\includegraphics{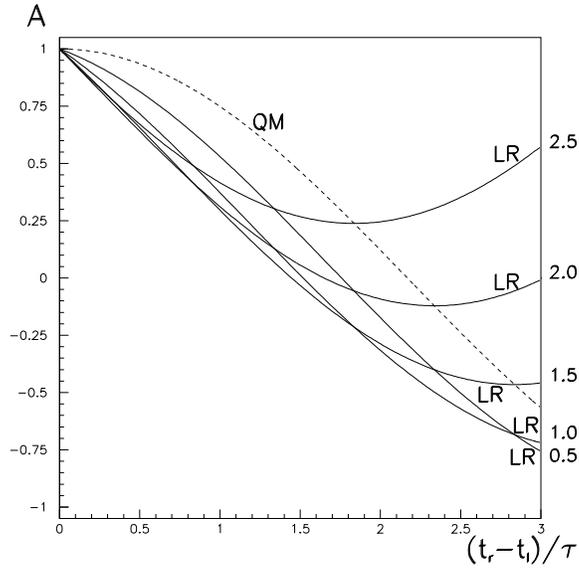}}
\end{center}
\caption{\label{fig2}Flavour asymmetry predicted by QM and LR (maximum values) for some fixed $t_{l}/\tau$ values.}
\end{figure}

\clearpage

\begin{figure}[ht]
\begin{tabular}{cc} 
\scalebox{.38}{\includegraphics{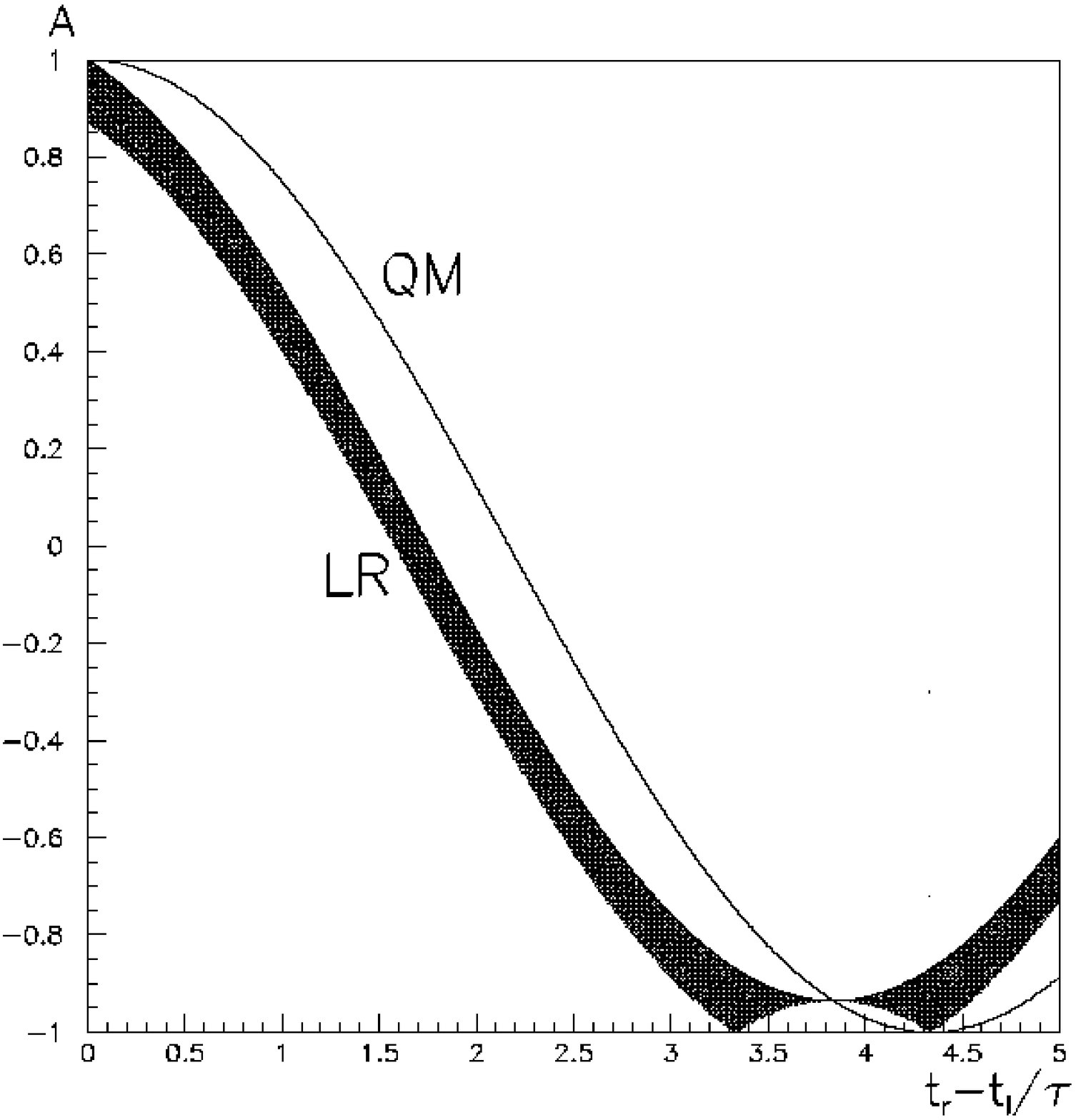}} &
\scalebox{.38}{\includegraphics{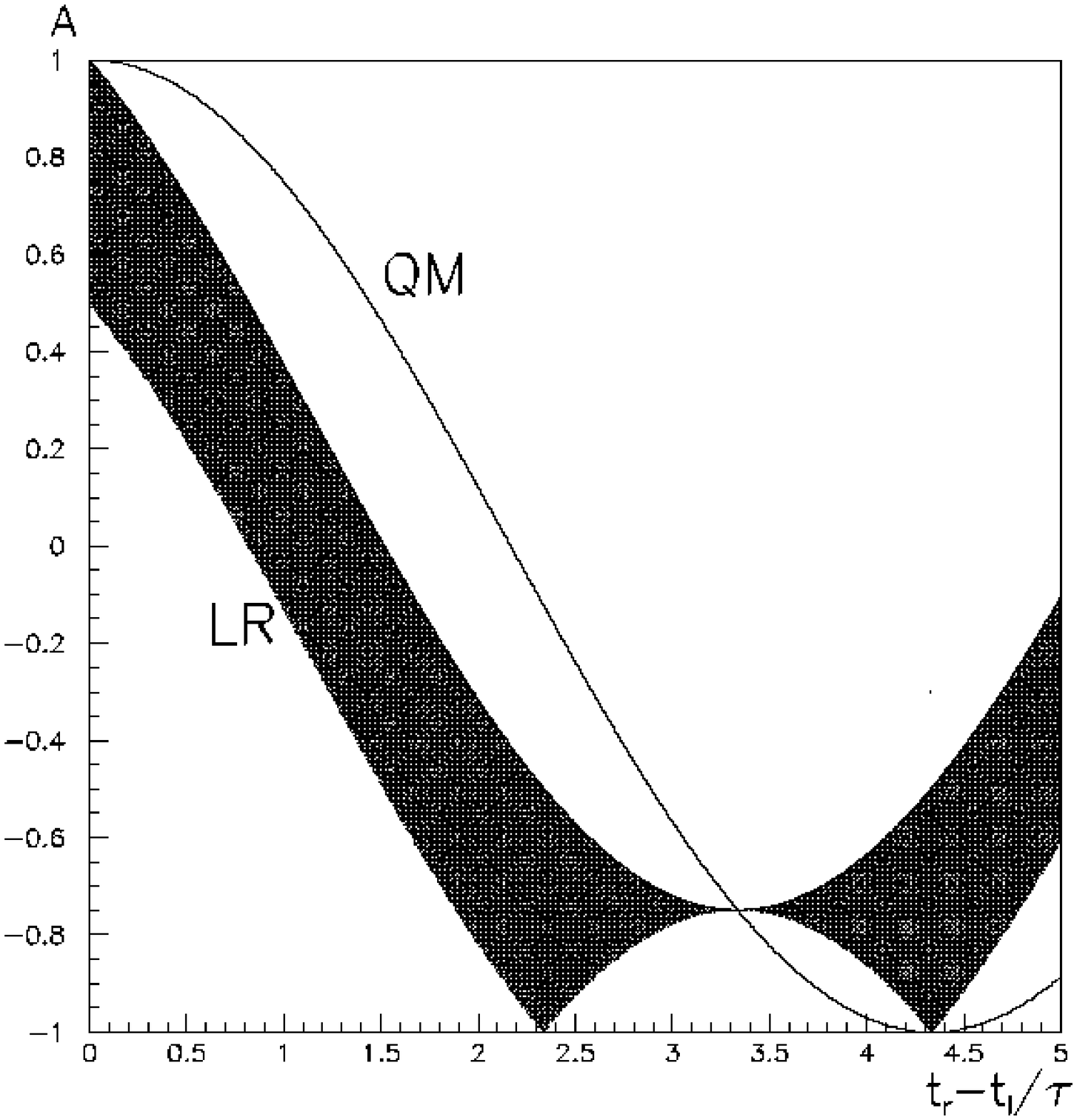}} \\
\end{tabular}
\caption{\label{fig3}Flavour asymmetry predicted by QM and LR for $t_{l}/\tau \! = \! 0.5$ (left) and $1.0$ (right).}
\end{figure}

\end{document}